\documentclass{sig-alternate-05-2015}

\usepackage{epstopdf}

\usepackage[show]{notes-alt}

\usepackage{float}
\usepackage{url}
\usepackage{hyperref}
\usepackage[all]{hypcap}
\usepackage{tabto}
\usepackage{amsmath}
\usepackage{enumitem}
\usepackage[autostyle]{csquotes}
\usepackage{listings}
\usepackage{subfig}

\usepackage[numbers,square]{natbib}

\lstdefinelanguage{json}{
	basicstyle=\normalfont\ttfamily,
	showstringspaces=false,
	breaklines=true,
	frame=lines
}

\begin{document}

\CopyrightYear{2016} 
\setcopyright{acmcopyright}
\conferenceinfo{JCDL '16,}{June 19-23, 2016, Newark, NJ, USA}
\isbn{978-1-4503-4229-2/16/06}\acmPrice{\$15.00}
\doi{http://dx.doi.org/10.1145/2910896.2910902}

\hyphenation{Map-Reduce}
\hyphenation{opti-mi-za-tion}

\title{ArchiveSpark:\\Efficient Web Archive Access, Extraction and Derivation\titlenote{This work is partly funded by the European Research Council under ALEXANDRIA (ERC 339233)}}

\numberofauthors{3}
\author{
\alignauthor
Helge Holzmann\\
       \affaddr{L3S Research Center}\\
       \affaddr{Appelstr. 9a}\\
       \affaddr{30167 Hanover, Germany}\\
       \email{holzmann@L3S.de}
\alignauthor
Vinay Goel\\
       \affaddr{Internet Archive}\\
       \affaddr{300 Funston Avenue}\\
       \affaddr{San Francisco, CA 94118}\\
       \email{vinay@archive.org}
\alignauthor
Avishek Anand\\
      \affaddr{L3S Research Center}\\
      \affaddr{Appelstr. 9a}\\
      \affaddr{30167 Hanover, Germany}\\
      \email{anand@L3S.de}
}

\maketitle

\begin{abstract}

Web archives are a valuable resource for researchers of various disciplines. However, to use them as a scholarly source, researchers require a tool that provides efficient access to Web archive data for extraction and derivation of smaller datasets. Besides efficient access we identify five other objectives based on practical researcher needs such as ease of use, extensibility and reusability.

\begin{sloppypar}
Towards these objectives we propose ArchiveSpark, a framework for efficient, distributed Web archive processing that builds a research corpus by working on existing and standardized data formats commonly held by Web archiving institutions. Performance optimizations in ArchiveSpark, facilitated by the use of a widely available metadata index, result in significant speed-ups of data processing. Our benchmarks show that ArchiveSpark is faster than alternative approaches without depending on any additional data stores while improving usability by seamlessly integrating queries and derivations with external tools.
\end{sloppypar}

\end{abstract}

%
%




\section{Introduction}
\label{sec:introduction}

A significant portion of the record of our society exists exclusively on the Web. Web archives aim to capture and preserve this record. Today, a large number of libraries, universities, and cultural heritage organizations have Web archiving programs \citep{bailey_ndsa2014}, with a 2011 survey reporting 42 different Web archiving initiatives across 26 countries \citep{gomes_tpdl2011}. With greater availability of Web archives and increasing recognition of their importance, a growing number of historians, social and political scientists, and researchers from other disciplines see them as rich resources for their research \citep{alexandria_hockx2014}. However, as Web archives grow in scope and size, they present unique challenges for creating tools and access methods for researchers.

One of the fundamental tasks in using Web archives for research is \emph{corpora building}. This task involves the selection and filtering of subsets, grouping and aggregation of records of interest and the extraction and derivation of new data (cp. Sec.~\ref{sec:usecase}). Consequently, there is a need for a framework that provides this functionality for efficiently constructing corpora out of the original archived collection. However, only providing fast access to the underlying collection is not sufficient. The framework needs to tackle a number of objectives driven by practical requirements (s. Sec.~\ref{sec:objectives_efficiency}), like simplicity, expressiveness, extensibility and the ability to produce reusable, well-structured output.

We address these objectives by proposing ArchiveSpark, a framework for distributed Web archive processing based on \textit{Apache Spark} (s. Sec.~\ref{sec:archive_spark}). By developing a tool solely based on standard file formats, we achieve the distinct advantage of institutions being able to easily share and apply the corpus generation specification across different collections. Towards providing efficient access, ArchiveSpark makes use of a metadata index (CDX) that is widely used by other tools in the domain of Web archiving. The CDX provides a lightweight representation comprised of metadata from all records in an archive. We achieve efficiency of access by exploiting the CDX to select records of interest before accessing the original archived content from disk. We also deliver substantial speed-ups by using \emph{lightweight representations} of records to enhance performance of distributed operations, like grouping and aggregation, unlike existing approaches that operate on much larger raw inputs. More specifically, rather than starting with all archived records and stripping them down, we operate on lightweight representation of records from the CDX and iteratively extend it as needed. We consequently observe large improvements in efficiency as we are able to minimize expensive disk operations involved as the researcher modifies and refines her requirements.

We compare and contrast our system with two alternative approaches and perform benchmarks to show differences in speed for select scenarios (s. Sec.~\ref{sec:benchmarks}). The benchmarks show that ArchiveSpark is faster than a similar approach that does not make use of the metadata index in the selected scenarios, which we aim at. Also, depending on the task, ArchiveSpark is even faster than a method of filtering based on HBase, a distributed database system, without the space and time overhead of ingesting and storing the archived data into a database.

ArchiveSpark is fully open-source and contributions to extend its functionality are very much appreciated. For this reason, we provide convenient extension points and an architecture that makes it easy to apply third-party tools to create custom derivatives from Web archives as part of an ArchiveSpark job specification. The working source code with the functionality that we provide out-of-the-box is available for open access:\\
\url{https://github.com/helgeho/ArchiveSpark}

\section{Use Case}
\label{sec:usecase}

In order to use Web archives as a scholarly source for scientific research, a required first step in most cases is the extraction of a well-defined corpus to work with \citep{alexandria_hockx2014, alexandriaKeynoteBruegger}. Scientists typically focus on a temporal and/or a topical subset of the archived data within the scope of their research question. In the following example, we consider five steps to be taken by a political scientist who wants to analyze sentiments and reactions on the Web from a previous election cycle.

\textbf{Step 1:} The researcher would need to define and extract a specific Web collection related to her research. In this case, she would only need websites that were archived in the time period of interest. However, this time-based or longitudinal filter alone would result in too many candidate websites as most Web archives are not topically organized. Finding just election related websites from this candidate pool requires domain expertise and/or manual intervention. For that reason, it is useful to have this pool to be as small as possible to begin with.

\begin{sloppypar}
\textbf{Step 2:} Since the researcher needs to consider only text resources from websites for her sentiment analysis project, she would apply a filter on \textit{MIME types} to only select such resources. However, identifying these resources by their MIME type involves accessing and parsing the HTTP headers of the records in the archive, which is a low-level detail and needs to be abstracted away from the potentially non-technical researcher.
\end{sloppypar}

\textbf{Step 3:} Another required filter involves the \textit{HTTP status code} of a particular capture. The fact that a certain URL was captured at some point in time and is part of the archive does not necessarily mean there was a valid Web resource being served at the URL. The URL could have been the result of an invalid link or a dead URL that was valid at a previous time. As our example researcher would only be interested in successful URL fetches, she would need to filter for records with status code 200.

\textbf{Step 4:} At this point, the candidate set is still likely to be very large for manual analysis. The researcher might decide to only focus on websites that contain certain terms or a specific set of entities, e.g., the candidates of the election. While seeming straightforward, this content-based filter involves accessing the content of every candidate record, which in turn involves separating the headers from the response body, encoding the textual response to a string, parsing out raw text from HTML, and finally applying text processing tools, before filtering on the resulting values.

\textbf{Step 5:} Web archives typically contain multiple captures of a website for every time the website was crawled, regardless of whether it has changed or not. Therefore, our researcher might decide to pick only the latest captures of the candidate URLs. In order to apply this filter, all captures of the intermediate corpus need to be grouped by URLs and sorted by their capture times. These types of operations are very expensive when performed on the raw records that include the entire payload. By operating only on metadata records that contain the required fields, they can be made much more efficient. However, this implementation is not something that the researcher should necessarily be concerned with.

Filtering, selection, grouping and extraction steps, like the ones described, can be arbitrarily continued. Depending on the task at hand, it may be necessary to keep track of where a certain value was derived from. As an example, consider the case that the researcher deems entities extracted from the title text to have more value than those extracted from the body text on a page. Keeping track of this lineage is an essential way to document the collection building and derivation process and enable its comprehension and reproduction by other researchers. Therefore, it should be included in the output format to be used by the researcher in her further research process.

ArchiveSpark seeks to tackle the challenges that arise by a research scenario such as the one described above. A researcher or a technical person supporting her on the corpus building process should be able to easily specify her requirements and efficiently extract the required corpus from a Web archive.

\section{Related Work}
\label{sec:related_work}

Scientifically published articles on data extraction from Web archives, like ArchiveSpark, have been very limited. To the best of our knowledge, the only comparable system is \textit{Warcbase} by \citet{warcbase}, which will be discussed at the end of this section and serves as the baseline in our benchmarking process (s. Sec~\ref{sec:benchmarks}). There are also a number of other approaches in the area of accessing and mining Web archives including tools from industry.

In this discussion on related work we differentiate between specialized Web archive access approaches based on certain properties and more general approaches. The former provide search and lookup operations as the method of access, while the latter provide access to all of the archived data with support for data processing. ArchiveSpark, our tool for general Web archive access, supports arbitrary filtering and data derivation operations on archived data making it much more suitable for the scientific use of Web archives.

\subsection{Specialized Web Archive Access}

The Internet Archive\footnote{\url{http://archive.org}}, one of the driving institutions of Web archiving, and most other Web archives, feature the \textit{Wayback Machine}\footnote{\url{https://github.com/iipc/openwayback}} to provide access to their Web collections. The Wayback Machine enables URL based access to the archived captures of a website, based on a server API powered by a metadata index (CDX). Lookups are designed for efficient, random URL based access and accomplished by running binary searches through the sorted index files. Researchers can query the CDX server for metadata information of a particular URL, host, domain or URL prefix.

In contrast to these structured queries by means of metadata, the \textit{UK Web Archive}\footnote{\url{http://www.webarchive.org.uk/}} is working on an information retrieval system based on the Apache Solr search platform\footnote{\url{https://github.com/ukwa/webarchive-discovery/wiki}}. Their \textit{Shine} project\footnote{\url{https://github.com/ukwa/shine/wiki}} supports faceted searching and more sophisticated trend analysis of Web archive content. \citet{alexandria_hockx2014} identifies 15 Web archives that feature similar kinds of full-text search capabilities. While these are largely engineering efforts that exploit existing search systems, there have also been scientific efforts to build indexes specialized on certain properties, such as time \citep{anand_sigir2011} or semantic annotations \citep{hinze_jcdl2015}. There are however two major challenges with these approaches that limit their applicability in the area of corpus building from Web archives. First, it is not always feasible to obtain the necessary resources to parse and index all archived Web content and store them in a search index. Second, even if the necessary resources are available, they cannot efficiently support corpus building processes that go beyond these specialized lookups. For these reasons, with ArchiveSpark, we propose a general data processing approach that exploits the CDX for gains in efficiency while not having to rely on an external index.

\subsection{General Web Archive Access}
\label{sec:general_access}

Due to the size of Web archives, often in the order of multiple terabytes, a single machine can no longer process or even store those collections. As a result, distributed computing facilities are commonly implemented for processing archived data. In contrast to the previously discussed specialized access approaches, these facilities enable general access to the archives by operating directly on the data records for selection, filtering, aggregation and transformation.

\begin{sloppypar}
As part of their self-guided workshops, like the \textit{Web Archive Analysis Workshop}\footnote{\url{https://webarchive.jira.com/wiki/display/Iresearch/Web+Archive+Analysis+Workshop}} and \textit{ARS Workshop} \footnote{\url{https://github.com/vinaygoel/ars-workshop}}, the Internet Archive provides a number of tools for this purpose. These tools enable researchers to batch process data and derive information like hyperlink graphs and mined text using, \textit{Apache Hadoop}, an open-source implementation of the MapReduce programming model for distributed computing of large datasets \citep{mapreduce}.
\end{sloppypar}

\citet{phd_alsum2014} presents with \textit{ArcContent} a tool for archive access based on Hadoop that uses Cassandra \citep{cassandra}, a distributed database to store the extracted data. Similar to ArchiveSpark, it involves a data filtering step where records of interest are selected using the metadata fields in the corresponding CDX dataset. However, in contrast to our approach, the extracted records are stored into Cassandra to be queried through APIs powered by a web service. This only works in cases where the research task is clear and well-defined beforehand and does not involve iterative filtering and data transformations.

Most similar to our ArchiveSpark framework is \textit{Warcbase} by \citet{warcbase}, an open-source platform for data processing on Web archives. It provides two different methods to access the data and serve as a baseline in our benchmarks (s. Sec.~\ref{sec:benchmarks}). Warcbase was originally developed to be based on \textit{HBase}, an open-source implementation of Google's Bigtable \citep{bigtable}, a Hadoop-based distributed database system. It features tools to ingest the Web archive records into HBase and allows for temporal browsing of URLs, with efficient, random URL based access similar to the Wayback Machine. The first method requires the storing of data in HBase with researchers leveraging Hadoop based tools to analyze it. However, this has the major drawback of involving an expensive setup phase of duplicating the entire Web archive in HBase. For the second method, Warcbase provides convenience functions to load and process the archive files directly using \textit{Apache Spark}, one of the most popular alternatives to Hadoop. Spark, in contrast to Hadoop/MapReduce makes extensive use of the main memory of nodes, which has shown to lead to impressive speed-ups \citep{spark}. On the GitHub repository of Warcbase, the authors recommend the Spark based method in order to avoid the HBase overhead of the first \citep{warcbase_github}. However, this Spark based method, in contrast to ArchiveSpark which is also based on Spark, does not optimize for efficiency or meet all of the objectives outlined below.

\section{Objectives}
\label{sec:objectives}

ArchiveSpark addresses six objectives, which we identified as being essential for a tool for corpus creation on Web archives, based on practical requirements. These comprise (1) a simple and expressive interface, (2) compliance to and reuse of the standard formats in the domain of Web archives, (3) an efficient selection and filtering process, (4) an easily extensible architecture to support various derivation tools, (5) lineage support to comprehend and reconstruct the process of derivation from the archive, and (6) an output in a standard, readable and reusable format.

\subsection{Simple and Expressive Interface}
\label{sec:objectives_interface}

The primary objective, when we designed ArchiveSpark, was a simple interface that lets users access the fields of interest without the need to do any parsing of archived Web records themselves. Users of this interface would be able to easily express any selection and filtering operations and access available information without carrying over complete archived records at each stage of the workflow. Additionally, the idea was to provide a seamless transition from filtering based on just metadata available in the index to that based on the contents of the archive.

Since ArchiveSpark is based on Spark (s. Section~\ref{sec:general_access}), which is written in Scala, we naturally chose Scala to be the language of choice for ArchiveSpark. Scala enabled us to specify the ArchiveSpark extraction and derivation workflow in a functional manner. This functional approach is less verbose than that of traditional object oriented languages and often simplifies tasks as it allows for a more natural way of expressing thoughts. Our interface is inspired by the existing Spark API and the Scala standard library, to provide the same degree of simplicity and expressiveness.

Even though the interface, in our opinion, is fairly intuitive to use by a computer scientist or a researcher familiar with programming, we do not expect researchers from other disciplines to be able to use it directly in all cases. However, with the aid of a technically savvy person, the researcher should be able to express her thoughts and requirements on the collection building process and get them easily translated into an executable ArchiveSpark workflow.

\subsection{Standard Formats}
\label{sec:objectives_formats}

In the area of Web archiving, there are a couple of file formats for storing archived web resources and derived metadata that have been established and in wide use over the years. As a result, these formats have either become de-facto standards or have been standardized by ISO. Given their common availability in almost every known Web archive, we wanted our system to be based on these file formats. We did not want to introduce any new file format or index structure: while such files or indexes could provide gains in efficiency for access, their generation would necessitate a pre-processing phase consuming expensive compute resources and additional storage. While being based purely on pre-existing file formats, ArchiveSpark maintains its essential objective of efficiency as described in the next sub-section.

The most important format in the world of Web archives is \textbf{WARC} (Web ARChive), which is registered as ISO 28500. WARC is a format to store archived web resources. Every record in a WARC file represents the capture of a single web resource at a given instant of time. The WARC record comprises a header section that includes the URL of the resource, the timestamp of capture and other metadata, as well as a payload section that contains the body returned by the web server. In the case of HTTP responses, the payload consists of a HTTP header and body. Before WARC was introduced as a format to store Web archives, archived records were widely stored in the older \textbf{ARC} format\footnote{\url{http://archive.org/web/researcher/ArcFileFormat.php}}. Although ARC is not standardized, many Web archives still contain data in this format, and hence ArchiveSpark supports both WARC and ARC file formats.

Another format which is not standardized but is seen as a de-facto standard is \textbf{CDX}\footnote{\url{http://archive.org/web/researcher/cdx_file_format.php}}. This is an index format that contains a number of metadata fields for every web capture including pointers to the (W)ARC file and the file-offset into the file where the capture is stored. A header line specifies the metadata fields contains in the plain text index file. Most commonly generated, however, are CDX files with either 9 or 11 fields, which are utilized by the Wayback Machine to serve records to users browsing the archive. Since the Wayback Machine software is currently the access method of choice for most Web archives, CDX files are generated by and/or readily available to these archives. As an example, CDX files are available for the crawls provided by the Common Crawl initiative\footnote{\url{http://blog.commoncrawl.org/2015/04/announcing-the-common-crawl-index}}. Furthermore, it is possible to generate both WARC and CDX files with the current version of the Unix/GNU download tool \textit{Wget}\footnote{\url{https://www.gnu.org/software/wget}}.

In summary, with ArchiveSpark we designed for, first, being compliant to these standard formats, and second, not introducing and depending on any new format. This way we aim to guarantee that any Web archiving institution that has (W)ARC and corresponding CDX files can use ArchiveSpark to extract and mine their Web collections, without requiring any expensive pre-processing steps or prerequisites.

\subsection{Efficiency}
\label{sec:objectives_efficiency}

Efficiency is one of the core objectives of ArchiveSpark. Since Web archives are typically large scale data collections of terabytes or even petabytes, a scan-based selection over all archive files is a very time consuming process and can potentially run in the order of multiple days. This is in most cases too inefficient to be used for corpus building as part of a scientific research task.

With ArchiveSpark we leverage the available CDX index files (s. Sec.~\ref{sec:objectives_formats}). As a first step, we apply filters on the metadata fields from the CDX and generate a small candidate pool with the captures of interest that need to be read in from (W)ARC files. This way, we potentially avoid the scenario of reading in all the records in the archive before ending up rejecting a large number of them (s. Sec~\ref{sec:approach}). Our approach of CDX-enabled filtering and selective data access results in efficiency gains over the scan-based approach.

Furthermore, when working with the raw archive records, complex operations, like groupings and aggregations, become much more expensive, since the whole records need to be moved around in a distributed setting. This could be optimized by stripping out data that is not required by those operations. However, if needed later, it will need to be recovered from disk, which is often even more expensive. 

With ArchiveSpark we turned this around using a selective data access and derivation approach, starting with lightweight records comprising of only metadata and iteratively extending them as needed, resulting in further gains in efficiency.

\subsection{Extensibility}
\label{sec:objectives_extensibility}

In most research applications, instead of working on the raw archived resources, a researcher is interested in extracting or deriving the data of interest for a given research task. Derivations can either be created from the original payload of an archived resource or from previously derived data. An example of such successive derivations on text are Natural Language Processing (NLP) tasks, such as the extraction of named entities from websites. The corresponding derivation tools operate on natural text and thus, first require the HTML parsers to remove markup and extract plain text, followed by the NLP tool, i.e., the named entity extractor, to extract the desired information.

There are a limitless number of other derivations that researchers can be interested in, e.g., audio/video fingerprinting on archived media files, OCR on archived images and many others. With ArchiveSpark we want to ensure any possible derivation from Web captures, regardless of whether they were constructed by us before-hand or not. Therefore, we designed a very flexible architecture with appropriate extension points that allow the application of custom code as well as third-party libraries to build derivatives from the records of a researcher corpus.

\subsection{Traceability}
\label{sec:objectives_lineage}

An important trait of any scholarly resource is transparency and traceability. In order to make scientific research reproducible it is essential to understand how the research corpus was designed. However, in the case of Web archives, it is difficult to retrospectively reproduce the crawling process. Reasons for this are, among others, an ever-changing Web, a semi-automatic prioritization by Web crawlers, changing crawling strategies as well as multiple, disparate parties being involved in the collection process. As a result, we found it even more important to focus on documenting the data lineage of corpus building from Web archives.

Also, depending on the needs of the researcher, it may often be sufficient to only deal with derived information and not include the original records. In order to reproduce this derivation process at a later time, a proper documentation of the data lineage is absolutely crucial. ArchiveSpark achieves this objective of traceability by documenting the data lineage of all the derived records. The documentation includes metadata that allows for the identification of all the source records responsible for the derivative as well as the the derivation path outlining the steps undertaken to filter, transform and derive from these records.

\subsection{Reusable Output}
\label{sec:objectives_output}

\begin{sloppypar}
The extraction and derivations steps performed by ArchiveSpark act as a preprocessing phase in a research pipeline. The data extracted from the Web archive serves as scholarly source for a research tasks, which can either be manual or programmatic. In the case of manual research, researchers would typically create rather small, very selective corpora and read in the results manually. On the other hand, researchers may use tools to analyze the corpora based on different features in a completely automatic or semi-automatic manner.
\end{sloppypar}

In either case, the corpus needs to be clean, well-structured and readable. While human readability implies a pretty printed output without too much clutter, machine readability implies data parsing support. The latter can be guaranteed best by producing data in a commonly used format with existing parsers for various programming languages. One such format is JSON, which was originally introduced as an exchange format for JavaScript to be used by Web services. However, because of its simplicity, it has become a widely used format that can be easily parsed by many pre-existing tools.

JSON supports a cascading nested structure with multiple levels of data and is therefore well-suited for supporting the data lineage functionality of ArchiveSpark (s. Sec~\ref{sec:objectives_lineage}). Another advantage is that these nested cascades of data can be easily presented in a fairly human readable form. For these reasons, we decided on JSON as the default output format of choice. Of course, any other output format that meets our outlined objectives can also be implemented and integrated into ArchiveSpark.

It is worth noting that the use of ArchiveSpark is not restricted to such an output. Researchers can also use it to access the archive, apply filters and derivations, and continue using the rich data types provided by ArchiveSpark in a Spark job to perform data analysis at scale, e.g., machine learning or graph analysis.

\section{ArchiveSpark}
\label{sec:archive_spark}

ArchiveSpark is a framework that enables efficient data access, extraction and derivation on Web archive data with a simple API that enables flexible and expressive queries. The following sections describe the approach as well as the distinct features of ArchiveSpark, which are designed to meet the previously described objectives.

\begin{figure*}
	\centering
	\includegraphics[width=0.85\textwidth]{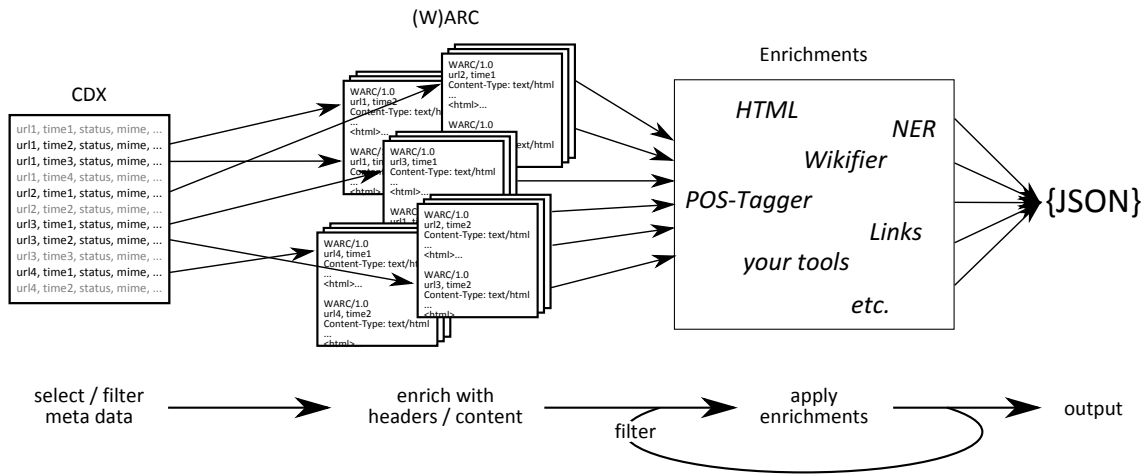}
	\vspace{5mm}
	\caption{Illustration of the ArchiveSpark selection and enrichment approach.}
	\label{fig:approach}
\end{figure*}

\subsection{Approach}
\label{sec:approach}

ArchiveSpark makes use of the CDX metadata index (s. Sec~\ref{sec:objectives_formats}) to selectively access resources from a Web archive. This approach is optimized for efficiency when extracting a defined subset of records as it avoids having to perform a full scan through all records in (W)ARC files. Since corpora used in scientific fields typically comprise of data derived from a small subset of the entire Web archive, ArchiveSpark is well suited for these use cases.

Figure~\ref{fig:approach} shows how ArchiveSpark works. First, the filtering process is performed using only metadata contained in the CDX files (s. Sec.~\ref{sec:objectives_formats}). Second, by utilizing the file pointers contained in the CDX records, ArchiveSpark selectively accesses the filtered records from the underlying (W)ARC files. At this stage, we augment the record's metadata with headers and content from the (W)ARC records. Next, users apply what we term \textit{enrichments} to derive new information, such as named entities or hyperlink data, that is added to the records. These enrichments can be applied by executing custom code or external tools. Based on the derived information, further filters and enrichments may be applied iteratively. The resulting corpus can be saved in a custom JSON format that is tailored to support data lineage.

\subsection{Interface}
\label{sec:interface}

The interface of ArchiveSpark is an API (Application Programming Interface) designed to define the specification of a Web archive extraction and derivation workflow. It is based on Apache Spark and greatly inspired by its API. Also ArchiveSpark uses the data structures of Spark and is hence fully compatible with any transformation methods provided by Spark. Like Spark, ArchiveSpark is implemented in Scala, a functional and object-oriented programming language running inside the JVM, Java's runtime environment. As a result, it is compatible with any third-party library running on the JVM as well, for instance all available Java and Scala libraries.

The entry point to ArchiveSpark is a globally available object with the same name. It serves as a starting point by providing methods to load Web archive files into so-called Spark RDDs (Resilient Distributed Datasets). RDDs are partitioned collections of objects spread across a cluster, stored in memory or on disk. Spark programs are written in terms of operations on RDDs.

Currently, we support reading in (W)ARC and CDX files that are stored in Hadoop HDFS (Hadoop Distributed File System). In order to load an ArchiveSpark RDD from HDFS, one simply needs to specify the path to the (W)ARC and corresponding CDX files. The following code is written in Scala, since it is our language of choice for defining an ArchiveSpark workflow specification:

\begin{lstlisting}
val archive = ArchiveSpark.hdfs(
	"/path/to/(W)ARC", "path/to/CDX")
\end{lstlisting}

The above \textit{archive} variable now references a Spark RDD consisting of specialized ArchiveSpark records. Hence, all methods provided by Spark to manipulate it through a set of parallel transformations, e.g., filter, as well as actions, e.g., count, can be applied. However, at this point these are based on the CDX data and therefore, only allow access to the metadata fields available in the CDX.

The following call applies filters on HTTP status codes and MIME types and only retains those records with a successful response (\textit{HTTP status code 200}) of type \textit{text/html}:

\begin{lstlisting}
val filtered = archive.filter(r =>
	r.status == 200 && r.mime == "text/html")
\end{lstlisting}

In the functional paradigm of Scala, every operation returns a new, immutable object instead of modifying the previous one. We have made sure this behavior is provided by ArchiveSpark as well. Hence, \textit{archive} still represents the entire dataset, while \textit{filtered} is a new object representing the filtered one.

As all Spark transformation operations are lazily evaluated, no actual data access will have been performed yet. The original RDD as well as the filtered one are just representations of the corpus to be extracted from the Web archive. The above filter is only evaluated or executed once a Spark \textit{action}, such as a data output, is performed. The advantage of the lazy loading is that, although all CDX records need to be read, only those that have passed through the filters are kept in the dataset consuming much less memory.

To access the actual content of these records in the next step, ArchiveSpark provides a method on archive record RDDs to apply so-called \textit{enrich functions}. The most basic enrich function is \textit{Response}. It opens the (W)ARC records, which are pointed to by the selected CDX records in the dataset, parses the HTTP response and enriches the original records with three fields: 1. (W)ARC header, 2. HTTP header, and 3. Payload:

\begin{lstlisting}
val response = filtered.enrich(Response)
\end{lstlisting}

Enrich functions can depend on each other and be applied consecutively. Each consecutive application derives new information from its parent dependency. While \textit{Response} does not depend on any other enrich function and is usually applied first, \textit{StringContent} depends on \textit{Response}. It transforms the payload of every record in the dataset into a string representation and enriches the record with this string. This works because our filter on the MIME types before made sure that our example dataset only contained text responses and no images or binary files:

\begin{lstlisting}
val strings = response.enrich(StringContent)
\end{lstlisting}

By explicitly enriching the records with both \textit{Response} and \textit{StringContent}, ArchiveSpark marks both these fields to be contained in the output. This way, by specifying what the records should be enriched with, the researcher can control the required features in the final corpus. If the dataset referenced by the \textit{response} variable had been directly enriched with \textit{StringContent}, only this enrichment would have been part of the output. However, internally, this process would still have first enriched the dataset with \textit{Response} as it is dependent on the payload. And since the payload was already present in the records from an earlier enrichment, the payload would have been used as-is and would not have needed to be re-computed. Note that dependencies specified in enrich functions are defaults but can also be explicitly specified by the user. For the sake of clarity and brevity, we do not show all the currently available methods and options of ArchiveSpark here in this paper.

Based on the enriched information, additional filters can be applied. This process of enriching and filtering can be repeated as needed. For the most efficient execution, it is recommended to apply filters as early as possible i.e. as soon as the data to be filtered on is available. This guarantees that any expensive derivation is performed on as few records as needed. This is especially important for the very first enrichment operation, which involves accessing data from (W)ARC files.

Other than the metadata fields available from the CDX records, the data derived by enrich functions is not typed, as different functions can create fields of various data types. The access to these values is enabled by specifying a path in dot-notation, where each segment specifies a level in the derivation pipeline. ArchiveSpark's \textit{get} method utilizes the ability of Scala to automatically infer data types based on their usage and casts the retrieved value into this type. As an example, the following instruction filters on the content string, i.e., the HTML code in the case of a webpage, and retains only those records that include the term \textit{internet}:

\begin{lstlisting}
val internet = strings.filter(
	r => r.get("payload.string").contains("internet"))
\end{lstlisting}

After the final dataset has been created, it can be written out as JSON using the \textit{saveAsJson} method on archive records RDDs provided by ArchiveSpark. It transforms the records into JSON objects consisting of the metadata and all explicitly enriched data:

\begin{lstlisting}
internet.saveAsJson("/output/path/results.json.gz")
\end{lstlisting}

The \textit{gz} extension is automatically detected by ArchiveSpark and causes it to compress the output using \textit{gzip}. The above six instructions have now created a corpus consisting of all successful text/html responses, i.e., HTML webpages, that contain the term \textit{internet}, formatted as pretty-printed and well-structured JSON in a compressed form. With this workflow approach, we believe we have met our objective of Section~\ref{sec:objectives_interface} of a simple and expressive interface.

As an alternative to the JSON output, users are free to transform the archive records that ArchiveSpark uses as its first class citizen into any form they want. We provide all the necessary access methods for this purpose. That way, besides the corpus building use case, ArchiveSpark can be used as a library to access Web archives as part of a larger data analysis application pipeline.

\subsection{Extensibility}
\label{sec:extensibility}

Currently, we provide the most basic enrich functions to get users started, but we will continue to extend ArchiveSpark with more functions moving forward. As ArchiveSpark is fully open source, any interested parties can also contribute to its development and provide their own tools as enrich functions. To support this, we provide convenient base functions that make it easy for a developer to define custom enrich functions meeting our objective of Section~\ref{sec:objectives_extensibility}.

An enrich function consists of the following four properties, which are required in the definition:

\begin{enumerate}[noitemsep]
\item \textbf{Dependency} The enrich function that this function depends on, e.g., \textit{Response}.
\item \textbf{Dependency Field} The resulting field of the enrich function that serves as input/source for this function, e.g., \textit{payload}.
\item \textbf{Result Fields} The resulting fields of this enrich function, e.g., \textit{string}.
\item \textbf{Body} The actual definition of the enrich function, specifying how new data is derived, i.e., the result fields, based on the original record or its dependency. The body can either consist of custom code performing the derivation or call an external tool.
\end{enumerate}

For the sake of simplicity, in addition to the above described \textit{enrich} method we also provide a \textit{mapEnrich} method on archive records RDDs. It allows a user to define enrichments without creating a specialized enrich function. This is especially handy if the enrichment is only used once, a very simple function or a highly custom one that is not worth the overhead of creating a new function. As an example, consider a function to obtain the length of a content string:

\begin{lstlisting}
val enriched = rdd.mapEnrich[String, Int](
	"payload.string",
	"length",
	s => s.length)
\end{lstlisting}

The syntax of such a \textit{mapEnrich} method is similar to the syntax of Spark's \textit{map} method or the \textit{map} method on standard Scala collections. However, in contrast to \textit{map} functions that transform one value into another, it enriches the original record with the resulting value preserving all the metadata and previously derived information. In the above content length example, \textit{String} specifies the input data type and \textit{Int} the output data type. The first parameter denotes the path from where to load the input and the second parameter names the result field. Unlike custom enrich functions, \textit{mapEnrich} methods can only create one result field. Instead of specifying the input path with dot separated field names, one can also pass in a dependency enrich function and the dependency field name. The last parameter of the \textit{mapEnrich} method is the body, which derives the required information, the content length in this case, from the value stored in the input path. Applying this method on a dataset creates a new record for each record in the dataset with the result field nested under the input path containing the result value of the body.

\subsection{Formats and Lineage Support}
\label{sec:formats}

Input files required by ArchiveSpark are WARC or ARC files with their corresponding CDX index datasets (s. Sec~\ref{sec:objectives_formats}). Currently, we support one of the most common CDX formats that is in use by the Internet Archive's Wayback Machine. This format encodes eight metadata fields and three additional fields pointing to the (W)ARC file where the capture is stored along with file-offset and compressed length of the record. However, we can easily support additional CDX metadata as the format evolves in the future.

CDX is a space-separated plain text format with each line representing one record. A single header line at the top of a CDX file denotes the fields: \textit{SURT URL (Sort-friendly URI Reordering Transform)}, \textit{timestamp}, \textit{original URL}, \textit{MIME type}, \textit{HTTP status code}, \textit{content digest/SHA-1 checksum}, \textit{redirect URL} (or \texttt{-}), \textit{meta tags} (or \texttt{-}), \textit{(W)ARC record compressed length}, \textit{(W)ARC record file-offset}, \textit{(W)ARC filename}.

An example CDX line looks as follows:\\
\texttt{com,example)/jcdl 20160117113253\\http://example.com/jcdl text/html 200 RKMS6XLYED4G8\\POFQUIN37WDEWYLD9Z - - 12345 67890 archive.warc.gz}

For the output format we decided on JSON, a widely used format that meets our objective of Section~\ref{sec:objectives_output}. Each output JSON record includes a listing of all the metadata fields from the source CDX identifying the selected resource. If no enrichments are applied, this would be the final output for our example record:

\begin{lstlisting}[language=json]
{
  "record": {
    "surtUrl": "com,example)/jcdl",
    "timestamp": "2016-01-17T11:32:53.000+01:00",
    "originalUrl": "http://example.com/jcdl",
    "mime": "text/html",
    "status": 200,
    "digest": "RKMS6XLYED4G8POFQUIN37WDEWYLD9Z",
    "redirectUrl": "-",
    "meta": "-"
  }
}
\end{lstlisting}

Enrichments are added to these JSON objects as additional keys next to \textit{record}. In case the \textit{Response} enrich function is applied, as in our example from Section~\ref{sec:interface}, the (W)ARC headers, HTTP headers as well as the raw bytes of the payload will be added in:

\begin{lstlisting}[language=json]
{
  "record": {...},
  "recordHeader":{
    "subject-uri": "http://www.example.com/",
    "content-type": "text/html",
    "creation-date": "20160117113253",
    ...
  },
  "httpHeader": {
    "Date":"Sun, 17 Jan 2016 10:32:53 GMT",
    "Connection":"close",
    "Content-Type":"text/html",
    ...
  },
  "payload": "bytes(length: 2345)"
}
\end{lstlisting}

Any other enrich function that depends on a value produced by \textit{Response}, e.g., payload, will result in the output being added as a nested value. If, for instance, the dataset was enriched with \textit{StringContent}, which calls \textit{Response} implicitly as its dependency, the resulting JSON might look like this:

\begin{lstlisting}[language=json]
{
  "record": {...},
  "payload": {
    "string": "<html>...</html>"
  }
}
\end{lstlisting}

In this case, the record and HTTP headers are not included, since the user did not explicitly specify them to be part of the corpus. The payload, however, is required to document the lineage of the string (the string representation of the payload). This meets the traceability objective of Section~\ref{sec:objectives_lineage} as every derived value can be traced back through the cascades to its origin.

When the user is interested in both the original value as well its derivations, for instance, when \textit{mapEnrich} is called to enrich the dataset records with their string content lengths (s. Sec~\ref{sec:extensibility}), a special underscore key (\texttt{\_}) is introduced. The field with this key retains the original value, like in the following example:

\begin{lstlisting}[language=json]
{
  "record": {...},
  "payload": {
    "string": {
      "_": "<html>...</html>",
      "length": 2345
    }
  }
}
\end{lstlisting}

Other derivatives based on this string content would be placed next to the underscore, just like \textit{length}. In the same way, if the dataset was explicitly enriched with both \textit{Response} and \textit{StringContent}, the byte representation of the payload along with the header fields would have been placed next to \textit{string}.

Finally, we consider the example of deriving named entities from the titles using the HTML string representation. This example would involve a HTML parser, which depends on StringContent to enrich the dataset with the required title value nested under a HTML field, as well as a named entity extractor tool, which in turn depends on the title to create a set of named entities. The lineage path of this constructed example would look as follows: \\\textit{payload.string.html.title.entities}.

\section{Benchmarks}
\label{sec:benchmarks}

\begin{figure*}[t]
	\centering

	\captionsetup[subfigure]{oneside,margin={0.85cm,0cm}}

	\subfloat[]{
		\label{fig:benchmarks_one_url}
		\includegraphics[width=0.32\textwidth]{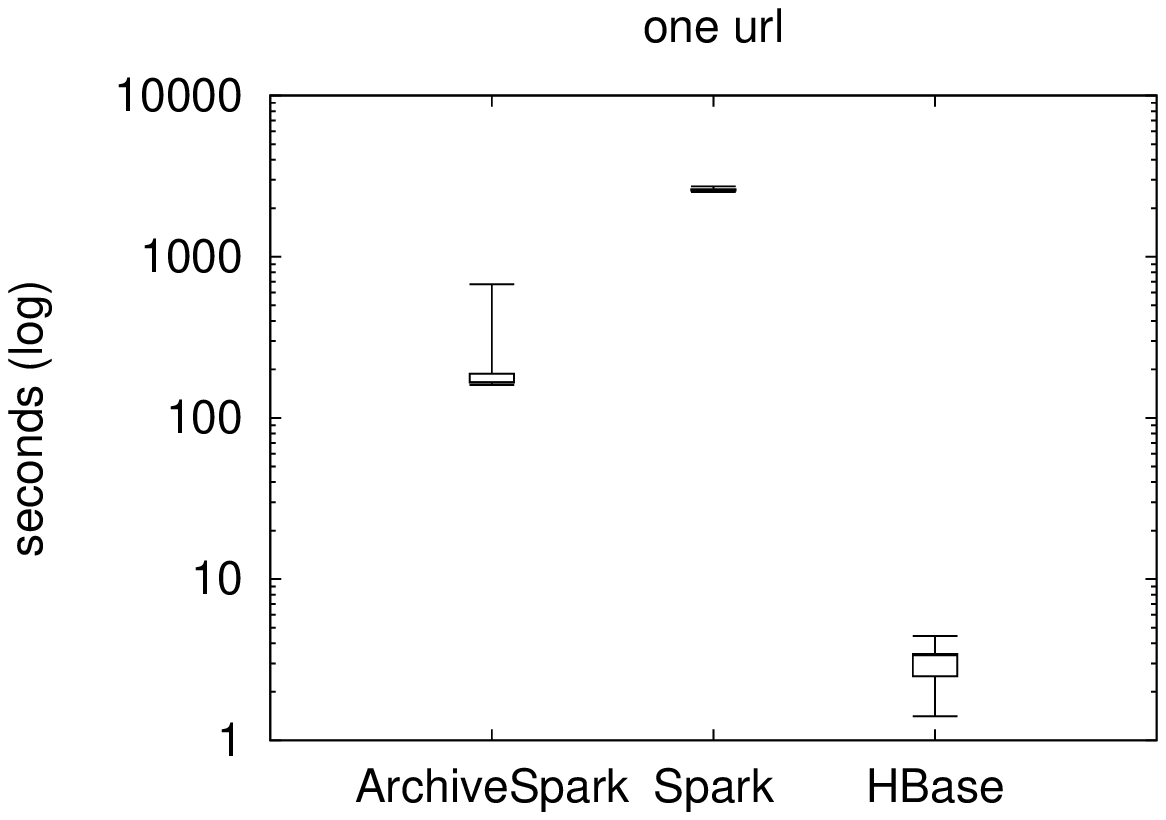}
	}
	\hspace{-1mm}
	\subfloat[]{
		\label{fig:benchmarks_one_domain_text_html}
		\includegraphics[width=0.32\textwidth]{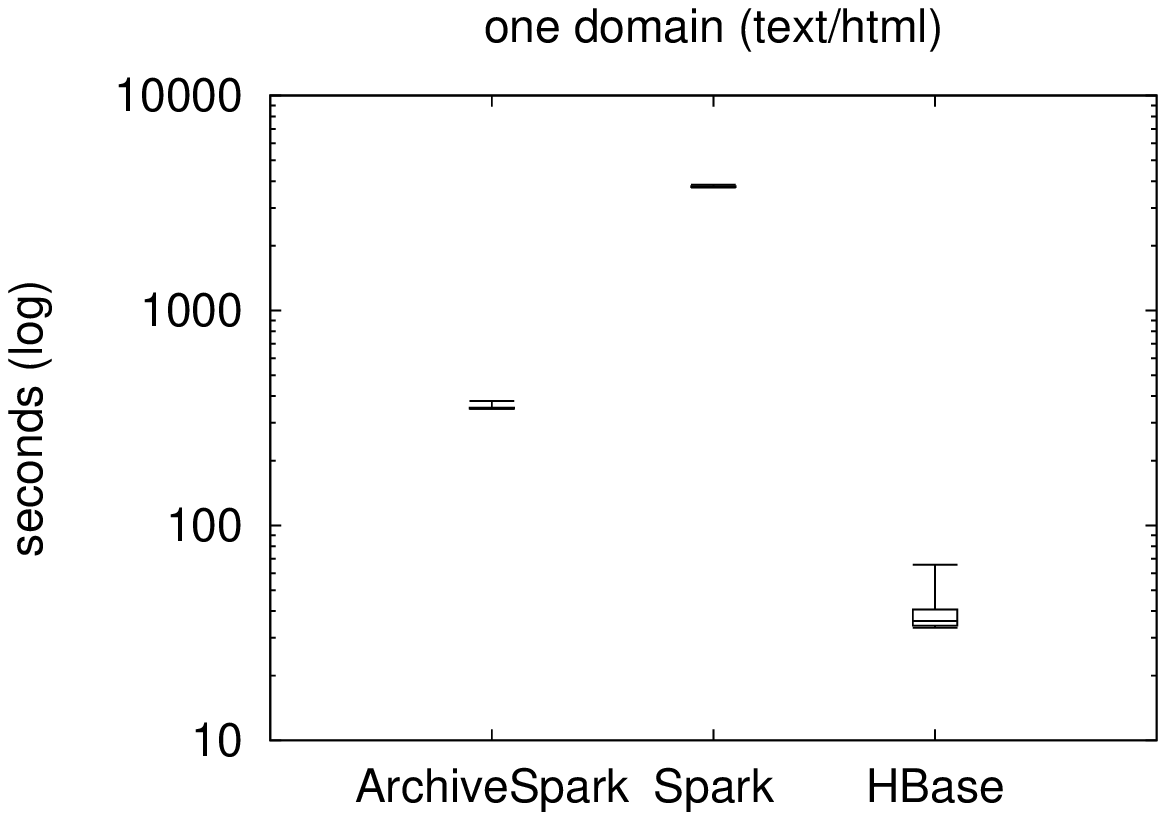}
	}
	\hspace{-1mm}
	\subfloat[]{
		\label{fig:benchmarks_one_month_latest_online}
		\includegraphics[width=0.32\textwidth]{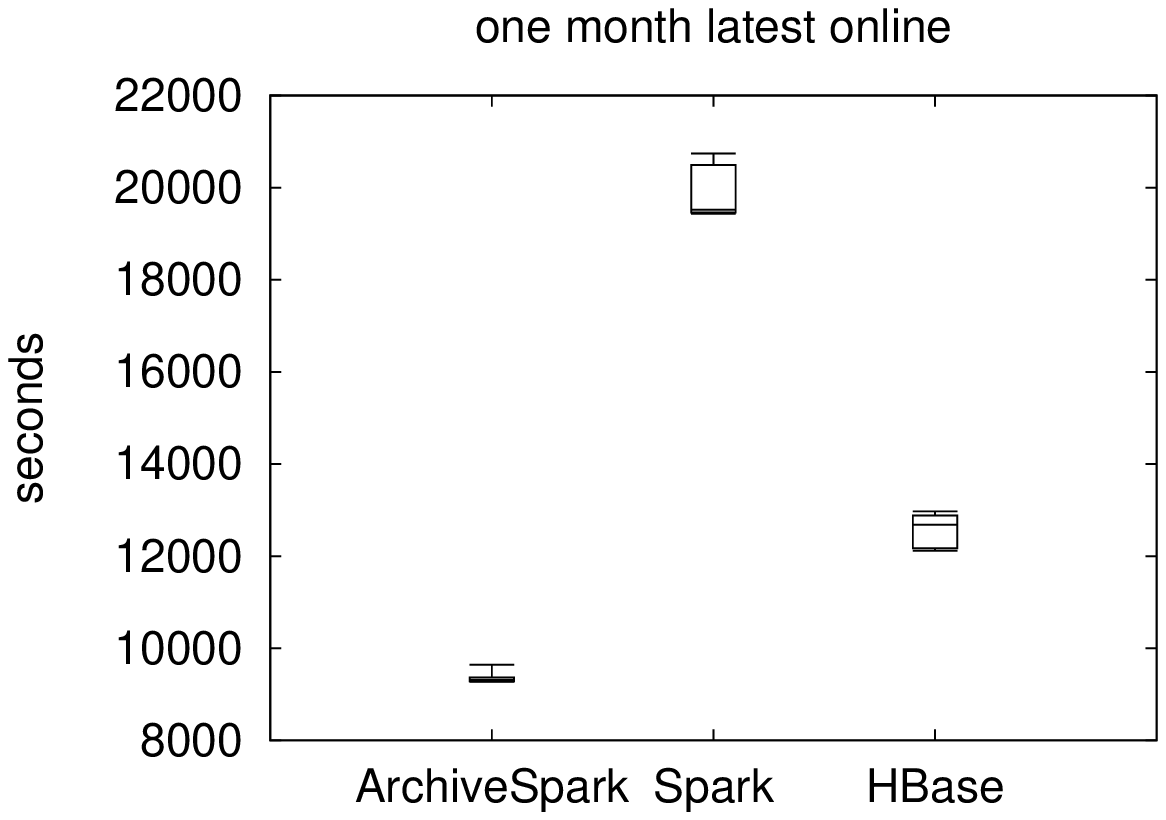}
	}

	\vspace{5mm}

	\caption{Benchmark times of ArchiveSpark vs. Spark vs. HBase (both by leveraging Warcbase)}
\end{figure*}

We ran benchmarks to assess the efficiency benefits of exploiting the CDX dataset when accessing Web archives (s. Sec.~\ref{sec:approach}). The run times of three different scenarios are compared using ArchiveSpark and two baseline approaches: a scan-based approach using pure Spark, and the Warcbase approach using HBase. For both baselines, we used the tools provided by Warcbase to load and access the datasets (cp. Sec.~\ref{sec:related_work}).

\subsection{Dataset}

One of the services provided by the Internet Archive is Archive-It\footnote{\url{https://archive-it.org}}. It is subscription based and enables partner institutions to run selective focused crawls to create and archive their own thematic and event driven collections. For our experiments, we chose one of these collections, the \textit{Occupy Movement 2011/2012}\footnote{\url{https://archive-it.org/collections/2950}} collection, collected by the Internet Archive itself. Unlike a generic Web crawl collection, this collection features a well-defined scope and is not too large, allowing our benchmarks to be performed in a reasonable amount of time.

The collection contains a total of 17,478,067 (17.4 Million) captures with 10,089,668 (10.08 Million) unique URLs. It contains Web content crawled during the time period Dec 3, 2011 to Oct 9, 2012, with a total storage of 470.9 GB of compressed WARC files. The CDX data, generated by us, adds in 24.4 GB of data size.

\subsection{Experimental Setup}
\label{sec:experimental_setup}

The experiments were performed on a Hadoop cluster running the Cloudera distribution\footnote{\url{http://www.cloudera.com}} (Hadoop 2.6.0-cdh5.4.9). The cluster consisted of 2 master nodes and 24 compute nodes with a total of 256 CPU cores, 2560 GB of RAM and 960 TB of hard disk space.

The three systems we compared in the benchmarks were:
\begin{enumerate}[noitemsep]
\item ArchiveSpark
\item Spark: Using Warcbase's Spark library
\item HBase: Using Warcbase's ingestion tool
\end{enumerate}
For both ArchiveSpark and pure Spark approach, WARC files from the collection were stored in Hadoop HDFS. The CDX files required by ArchiveSpark were generated using the Internet Archive's CDX Generator, which is available open source on GitHub\footnote{\url{https://github.com/internetarchive}}. Generating the CDX files took 110 minutes, however, this is a one time process and is anyway a necessary step to enable access services like the Wayback Machine. This dataset could have also been downloaded directly from Archive-It. For these reasons, we consider this CDX generation step to be negligible in the benchmarks.

In the HBase (Warcbase) approach, we had to first ingest WARC files into HBase. Warcbase exploits certain properties of HBase to enable access to Web archives. For instance, different captures of a crawled Web resource are stored as timestamped versions of the same record in HBase. URLs are stored in an inverted, sort-friendly format and are used as row keys for fast lookups with the MIME type serving as a column qualifier. These design decisions allow for an efficient selection and filtering process based on these three properties: URL, timestamp of capture, and the MIME type. When additional fields are required, those need to be parsed from the WARC records, either from headers or the payload, which are stored as values in HBase cells. Due to limitations on the local disk space of our cluster, we had to ingest the data into HBase from the WARC files stored in HDFS. As the current version of Warcbase only supports reading in WARC files from the local file system, we modified this system accordingly. The ingesting process took a little over 24 hours with the resulting database containing a complete copy of the entire collection.

For both the Spark and HBase approaches, we queried the data using Spark and also used it to perform operations on the resulting data. All three systems being compared ran with the same Spark configurations, using 10 executors with 4 GB of memory each. As the cluster was not exclusively available to us, with other jobs running at the same time, the cluster load varied among the benchmarks. To compensate for these variations, we ran every single benchmark a total of five times.

We chose a common task among all benchmarks: select a subset of records from the entire dataset, count the length of the string content of these records and compute the sum of these lengths. This task is well-suited for the benchmarking process since it features the extraction workflow supported by ArchiveSpark. It involves a filtering phase to select the subset of records of interest, an enrichment phase to augment records with content, as well as a derivation phase that enriches the content with its string representation and length. We intentionally did not apply any more sophisticated enrichments that involved third-party libraries as those would only be applied on top of these results and would depend on the performance of these external tools.

\subsection{Scenarios and Results}

The benchmark consisted of three different scenarios, starting with the most basic filtering operation to only select records of a given URL, and ending with a more sophisticated scenario involving a grouping operation to select the latest online capture of all URL from a specific time period.

\subsubsection{Scenario 1}

First, we filtered the dataset for all records of one particular URL, i.e., \textit{http://map.15october.net/reports/view/590/}. In case of HBase, this is directly supported and constitutes a simple row query. Therefore, it is understandably very fast with the query taking between 1.4 and 4.4 seconds. However, when comparing with the other approaches, the pre-processing time required for HBase as well as the additional space requirements need to be kept in mind (s. Sec.~\ref{sec:experimental_setup}). The times of all three approaches are illustrated in Figure~\ref{fig:benchmarks_one_url}, where the whiskers represent the fastest and slowest runs, while the box covers the ones in the middle, with a centered line representing the median. As shown, ArchiveSpark is about 100 times slower with times between 160.3 and 675.4 seconds, but still around 10 times faster than pure Spark with times between 2522.6 and 2734.0 seconds. This is where ArchiveSpark's incorporation of the CDX index leads to performance benefits as it allows for the selective access of only records of the given URL, while pure Spark performs a scan over the entire dataset and parses every single record in order to find these records.

\subsubsection{Scenario 2}

In the second scenario, instead of filtering by URL, we selected all webpages, i.e., MIME type \textit{text/html}, belonging to a specific domain, i.e., \textit{15october.net}. The results are shown in Figure~\ref{fig:benchmarks_one_domain_text_html}. The HBase query performs a targeted row scan again, this time for all keys starting with the specified domain in its inverted, sort-friendly form, i.e., \textit{net.15october}). However, this alone is not sufficient as the scan would also yield rows starting with \textit{net.15octoberx}, which is not the correct domain. Therefore, an additional filtering step is required. Next, the filter by MIME type \textit{text/html} is also directly supported by HBase, since MIME type is available as a column label. With times between 33.4 and 65.6 seconds, the HBase approach is around a magnitude of 10 slower than in the first scenario. ArchiveSpark comes closer to HBase with times between 349.2 and 379.1 seconds, because both values to be filtered are part of the CDX and therefore, the task is similar to the one in the first scenario. The pure Spark approach of a complete scan is around 10 times slower than ArchiveSpark with times between 3737.7 and 3853.2 seconds.

\subsubsection{Scenario 3}

Finally, we selected the latest successful captures for all URLs crawled in a specific month, i.e., Dec 2011. This is accomplished in two steps: first, all captures from the desired time period (Dec 2011) and with a successful response (status code 200) are selected and next, the latest capture for each candidate URL is chosen. The pure Spark approach takes between 19432.0 and 20744.3 seconds in this scenario. This approach first scans through all records of the dataset, followed by the step of identifying the latest capture of every URL from the set of qualifying records. This may be more efficient when only a few records of a dataset need to be filtered out. However, in scenarios, like this, where users are interested in only a small subset of a large collection, it is very slow. In the HBase approach, although HBase directly supports timestamp based filtering, which is performed on the versions of a URL, filtering on the HTTP status code requires parsing the WARC record to read in the status code. Only then can the latest successful captures be selected as an additional post-processing step. The HBase approach takes between 12117.7 and 12971.5 seconds. For ArchiveSpark, as both properties, timestamp and HTTP status, are contained in the CDX files (cp. Sec.~\ref{sec:formats}), the filtering as well as selection of the latest captures is entirely possible using just the CDX. For that reason, ArchiveSpark leads in this benchmark as illustrated in Figure~\ref{fig:benchmarks_one_month_latest_online} with times between 9639.6 and 9270.8 seconds. This illustrates how the rich potential of ArchiveSpark's selective access approach is unlocked when a large fraction of the dataset can be filtered out based on available metadata.

\section{Conclusion and Outlook}
\label{sec:conclusion}

Web archives are becoming more and more important as a scholarly source and building a corpus from these archives is typically one of the first steps in any research process. Since researchers working with these Web collections are often from the humanities with no technical background, there is clearly a need to simplify this extraction and derivation process. In the first part of this paper, we presented a number of objectives and discussed why we deem them as essential for any system that supports building research corpora from Web archives. These include simplicity in terms of usage and extensibility, efficiency of access and traceability by documenting data lineage for the purposes of reproduction and reuse.

In the second half of the paper, we presented ArchiveSpark, a framework that effectively tackles these objectives by making use of existing file formats, a functional approach to data processing at scale and utilizing a widely deployed metadata index. By utilizing this index that is a de-facto standard in the area of Web archiving, ArchiveSpark avoids having to perform any pre-processing of the data or having to invest in additional storage space. We also provided benchmarks that show how ArchiveSpark is more efficient than other alternatives when selecting records of interest based on the rich metadata already available in the metadata index. ArchiveSpark, however, is not the best option when a data processing task needs to run across all or a large fraction of the records in a Web archive.

Moving forward, we plan to extend ArchiveSpark to support more data sources, such as streaming data over HTTP, which would allow researchers to efficiently extract corpora from publicly available, remote Web archives without needing a local copy of the complete dataset. Since Python is a popular language among data researchers and scientists, we plan to provide support for PySpark, the Python API for Apache Spark. ArchiveSpark is fully open source, and we hope for many contributions from the broader community, especially in terms of third-party tools to be used as extensions in the ArchiveSpark pipeline.

\renewcommand*{\bibfont}{\raggedright}
\bibliographystyle{unsrtnat}
\bibliography{references}

\end{document}